\documentclass[aps,amsmath,twocolumn]{revtex4}

\usepackage{graphicx}
\usepackage{dcolumn}
\usepackage{bm}
\usepackage{color}
\usepackage[normalem]{ulem}

\begin{document}

\title{Laser stabilization to an atomic transition\\ using
an optically generated dispersive lineshape}

\author{Fabiano Queiroga, Weliton Soares Martins, Valdeci Mestre, Itamar Vidal, Thierry Passerat de Silans, Marcos Ori\'{a}, and Martine Chevrollier}
  \affiliation{Laborat\'{o}rio de Espectroscopia \'{O}tica, DF-CCEN \\ Cx. Postal 5086 - Universidade Federal da Para\'{i}ba\\58051 - 900 Jo\~{a}o Pessoa - PB, BRAZIL\\
\email{martine@otica.ufpb.br}}%

\date{\today}
\begin{abstract}
We report on a simple and robust technique to generate a dispersive signal which serves as an error signal to electronically stabilize a monomode cw laser emitting around an atomic resonance. We explore nonlinear effects in the laser beam propagation through a resonant vapor by way of spatial filtering. The performance of this technique is validated by locking semiconductor lasers to the cesium and rubidium $D_2$ line and observing long-term reduction of the emission frequency drifts, making the laser well adapted for many atomic physics applications.
\end{abstract}
\maketitle
\section{Introduction}
\indent The use of lasers in atomic physics often demands long term stability of the central frequency of the light emission. For metrological applications the stabilization technique \cite{Hall1999,Salomon1988} should be very carefully chosen and applied, frequently controlling the laser linewidth and avoiding to introduce any artificial shift in the laser emission frequency. Moreover, locking to the center of an atomic or molecular transition usually requires modulation techniques and lock-in detection. On the other hand, for many scientific and technical applications one only needs to avoid frequency drifts and, sometimes, the desired laser frequency does not lay at the maximum of an atomic lineshape, but rather at a displaced frequency, as for instance, when operating an optical cooler \cite{Lett1989}. For such applications a few simple and reliable techniques were developed \cite{Cheron1994,Corwin1998,Pearman2002,Debs2008,Robins2002,Gazazyan2007,Martins2010} and allow on
 e to deal with lasers in various long run experiments.  The main idea behind many of these techniques is to generate a dispersive lineshape that will produce an error signal. In particular, for the dichroic atomic vapor laser lock (DAVLL) \cite{Cheron1994} and its variants \cite{Corwin1998,Martins2010}, the stabilization frequency may easily be chosen around the center of the Doppler-broadened line. However, a relatively uniform external magnetic field is needed to generate the Zeeman split of the probed hyperfine transition and a double detection with well balanced photodetectors is also necessary.\\
	
\begin{figure}[h!]
\centering
\includegraphics[width=7.5cm]{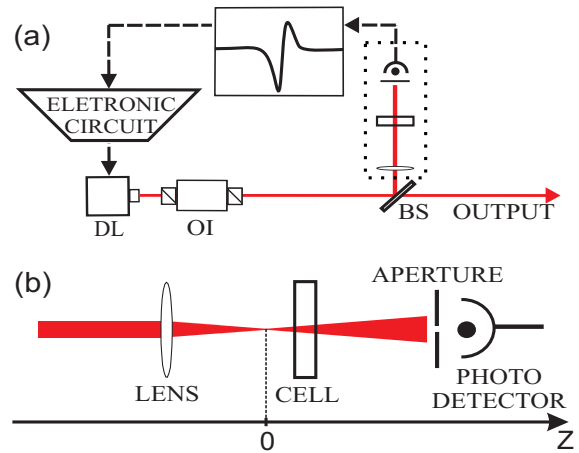}
\caption{(Color online) (a) Experimental setup. Part of the laser beam is used to generate a dispersive like curve that is fed to an electronic circuit used to control the laser frequency. DL, Diode laser. OI, Optical isolator. BS, Beam splitter. (b) Detail of the beam configuration for ANGELLS}
\label{setup}
\end{figure}

	\indent In this work we report on a simple method to generate a dispersive signal in a very direct way, and therefore of easy implementation. Our technique explores the dispersive signal obtained when a Gaussian-profile light beam is sent through an atomic vapor cell and is detected after spatial filtering by an aperture (Fig. \ref{setup}). We call this method ANGELLS, an acronym for Atomic Non-linearly GEnerated Laser Locking Signal.\\
\section{Origin of the dispersive lineshape}
	\indent The third order susceptibility term of an induced atomic vapor polarization by a laser beam results on a non-linear refractive index term, proportional to the laser intensity. The total refractive index of the vapor can thus be written as $n(I)=n_0+n_2I$ and the radial intensity gradient of a Gaussian-profile beam will induce a radial refractive index gradient in the medium. This index gradient will in turn act as a lens for the gaussian beam, which will therefore suffer (self)-focusing or (self)-defocusing, depending on the sign of the nonlinear refractive index \cite{selffocus}. The nonlinear index changes sign across a sharp resonance of the nonlinear medium. If on one side of the resonance frequency the index increment is positive (maximum on the beam axis), the medium behaves as a converging lens and the power of a initially collimated beam transmitted through an aperture will increase (peak of the dispersive lineshape). On the other side of the resonance frequency the laser-induced increment is negative (minimum on the axis), the medium behaves as a divergent lens and the transmission through an aperture yields a correspondingly diminished signal (valley of the dispersive lineshape). In other words, the nonlinear medium acts as a lens which focal length depends on the laser frequency. For a hot atomic vapor, for instance, the non-linear refractive index can be written as \cite{n2}:
	\begin{equation}
	n_2\propto\delta e^{-\delta^2/\delta_D^2},
	\end{equation}
where $\delta_D=\frac{k}{2\pi}\sqrt{2k_BT/M}$ is the Doppler width, $T$ is the vapor temperature, $k_B$ is the Boltzmann constant, $k$ is the light wavenumber, $M$ is the atomic mass and $\delta$ is the laser frequency detuning relative to the atomic transition. For red detuning frequencies ($\delta<0$), $n_2$ is negative while for blue detuning ($\delta>0$), $n_2$ is positive. The power transmitted through the aperture is thus modulated when the frequency is scanned around {\it an atomic transition}, resulting in a dispersive-like lineshape with {\it Doppler width}.   \\

\section{Experiment}
\begin{figure}[btp]
\centering
\includegraphics[width=8.0cm]{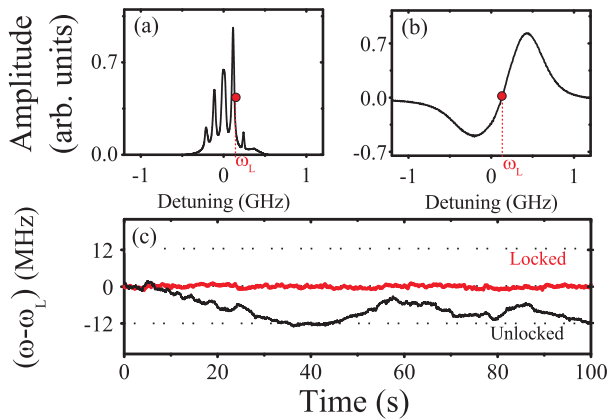}
\caption{(Color online) (a) Cs saturated absorption (SA) spectrum from an extra reference cell (not shown in Fig. \ref{setup}), (b) dispersion curve for the error signal as a function of frequency detuning relative to the atomic resonance, and (c) the ANGELLS-frequency-locked as well as an unlocked SA signal. The chosen frequency of stabilization ($\omega_L$) is marked by a red dot in spectra (a) and (b)}
\label{locked}
\end{figure}

\indent In our technique, the nonlinear medium is a resonant atomic vapor, placed in the laser beam path past a converging lens to enhance nonlinear effects with higher light intensity radial gradients. Our experimental setup is sketched in Fig. \ref{setup}. A 852 nm cw tunable semiconductor laser beam is splitted by a 90/10 beam splitter. The lower-intensity beam ($\sim50 \mu W$), of approximately Gaussian spatial profile (no spatial filter needed) and of slightly saturating intensity, is focused by a 150 mm-focal-length lens. A warm (40 to $\approx$ 60$^\circ$C, corresponding to densities of $2\times10^{11} at/cm^3$ - $10^{12} at/cm^3$) atomic cesium vapor \cite{cesium} contained in a 1-mm-long optical cell \cite{cell} is placed close ($\approx$ 20 mm) to the focus of the laser beam. We detect the transmission of the laser beam through an aperture adjusted so as to capture $\sim20\%$ of the beam power (typically 2-mm aperture for a beam of diameter 6 mm). When the frequency
  is scanned around the Cs $6 S_{1/2}, F=4 \longrightarrow 6 P_{3/2},F'=3,4,5$ Doppler transition, the nonlinear refraction turns from self-focusing to self-defocusing. This gives rise to a dispersive-like lineshape superimposed to a non-zero offset corresponding to the out-of-resonance aperture transmission (no vapor-induced modifications). Very small structures on these spectra are attributed to non-linear effects due to the beam reflection on the cell windows. The comparable dimensions of the beam diameter and the cell thickness makes the prevention of this high-order interaction difficult. However, this does not pose any additional problem to lock the laser at any position in the broad range inside the Doppler width. \\
\indent The error signal is the subtraction of a reference voltage (corresponding to a frequency within the Doppler line) from the photodetector amplifier voltage. Such an error signal is amplified and sent to the control of the laser frequency to correct for laser frequency drifts. In semiconductor lasers, the frequency is changed mostly through the injection current, the junction temperature or, in the case of extended cavity configuration, the external diffraction grating angle. We worked with a DFB laser diode resonant with the Cs D$_2$ line and a Fabry-P\'{e}rot semiconductor laser with extended cavity, emitting around the Rb D$_2$ line. The electronic correction signal is fed back in the junction current in the DFB or in the piezoelectric actuator in the extended-cavity laser \cite{DF}. For the sake of simplicity, we have operated both systems with a home-made electronic circuit having only proportional and integral gains.\\
\begin{figure}[htb]
\centering
\includegraphics[width=7.5cm]{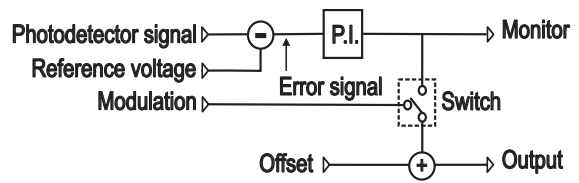}
\caption{Schematics of the circuit to choose and lock the laser to an atomic frequency.}
\label{circuit}
\end{figure}
\indent A scheme of the locking circuit is shown in Figure \ref{circuit}. A voltage ramp generator allows one to scan the laser frequency (through current or cavity PZT modulation) around the atomic resonance. We choose a locking frequency with the help of a reference saturated absorption (SA) spectrum carried out in an extra vapor cell and exhibiting characteristic sub-Doppler features (see Fig. \ref{locked}a). We use the SA signal obtained in this additional cell as frequency reference (Fig. \ref{locked}a) as well as to monitor the locking performance (Fig. \ref{locked}c). The locking procedure follows some basic steps: the ramp is turned off and the offset finely tuned until the laser frequency is at the desired locking point {\bf($\omega_L$)}, marked by dots in Figs. \ref{locked}a and \ref{locked}b. The error signal is then brought to zero by adjusting the reference voltage (Fig. \ref{locked}b) and a switch closes the loop, ultimately locking the laser at the desired frequency (Fig. \ref{locked}c). Modifying the reference voltage allows one to lock the laser at any point within the Doppler width and thus to explore the different hyperfine transitions shown in the SA spectrum. Notice that the reference voltage brings the error signal around zero and so compensates for the non-resonant background signal.\\ 
\section{Results}
\indent Figure \ref{locked}c exhibits the SA signal with the laser locked at the selected frequency, over a period of a few minutes, as well as the SA signal for an unlocked laser. The system remains locked for hours even after we have strongly and repeatedly hit our home-made optical table. The short-term rms frequency width is the same for the locked and the unlocked laser, i.e. of the order of 2 MHz or less, as measured using the saturated absorption line flank as a frequency discriminator. The long-term frequency fluctuations of the locked laser remains limited to less than 2 MHz rms, while the frequency of the laser unlocked for a few minutes fluctuates in excess of 20 MHz. \\ 
\begin{figure}[!h]
\centering
\includegraphics[width=8.0cm]{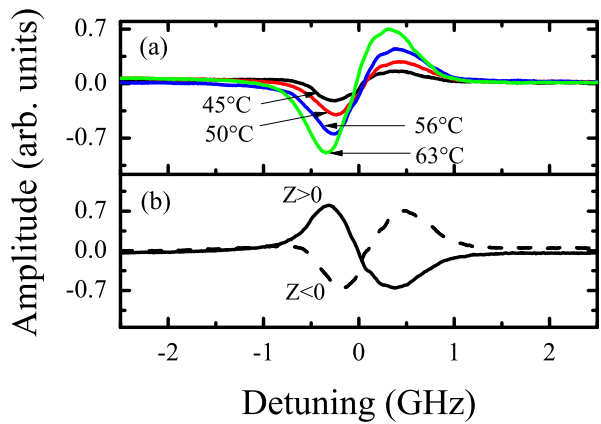}
\caption{(Color online) Error signal as a function of frequency detuning relative to atomic resonance (a) for different temperatures of the atomic vapor and (b) when the cell (Cs vapor) is placed a few millimeters before ($z<0$, dashed line) and after ($z>0$, solid line) the beam minimum waist.}
\label{dispersion}
\end{figure}
\indent We checked the stabilization sensitivity to vapor density and cell alignment along the beam. Although the vapor temperature has been varied between 45 and 63 $^\circ$C, the lineshape of the generated signal is stable against temperature changes as shown in Fig. \ref{dispersion}a. The Doppler-profile center position changes very little ($\approx 2$MHz/$^\circ$C) over this 20 $^\circ$C temperature variation. Each lineshape in Fig.  \ref{dispersion}a has been recorded at a given temperature (values indicated in the figure frame), fluctuating less than 1 $^\circ$C from its reference value. Thus, the laser frequency locking, particularly at the Doppler center, is not affected by small temperatures fluctuations of the vapor which lead to frequency drifts of the order of magnitude or less than the rms frequency width, as evidenced in Fig.\ref{locked}c. Even to lock the laser at frequencies other than at the line center, we have only monitored the cell temperatur
 e, without active control. More noticeable modifications appear on the profile wings, that do not play a role in the stabilization process inside the Doppler width. Similarly, the error signal remains approximately unchanged over $\approx$ 2-mm displacements along the beam, around the optimal position of the cell ($\approx$ 20 mm on either side from focal point). Another characteristic of using the ANGELLS technique is the possibility of choosing the sign of the error signal slope by purely optical means, as shown in Figure \ref{dispersion}b. The dispersive curve gets inverted when the vapor is displaced \textit{across the focal position} (see Fig.\ref{setup}). We also emphasize here the fact that semiconductor lasers are well known to exhibit very stable amplitude \cite{AmpNoise}, allowing us to disregard amplitude noise in the detected signal. For lasers with higher intensity fluctuations a second photodetector may be used to 'normalize' the frequency error signal.\\
 \section{Conclusion}
 
\indent In summary, we have presented an opto-electronic stabilization method based on the direct generation of an optical dispersive-like signal. The ANGELLS technique has proved to be an easy and robust locking method against diode laser frequency drifts. For similar performance, the set-up is simpler than  traditional locking techniques, not requiring magnetic fields or beam modulation. The set-up is flexible: different combinations of laser power, beam diameter, focalization, cell length and vapor density have been used, the values given in the article corresponding to those used for the presented results.

ACKNOWLEDGMENT: This work was partially funded by Conselho Nacional de Desenvolvimento Cient\'{i}fico e Tecnol\'{o}gico (CNPq, contract 472353/2009-8) and Coordena\c{c}\~{a}o de Aperfei\c{c}oamento de Pessoal de N\'{i}vel Superior (CAPES). F.Q., W.S.M., I.V. and M.C. acknowledge grants by CNPq.


\begin{thebibliography}{99}
\bibitem{cesium}We have performed similar experiments of laser stabilization using an atomic Rb vapor to lock a 780 nm diode laser to different transitions of the two rubidium isotopes.
\bibitem{cell}The cell length was chosen to have a thickness small compared to the lens focal length. Stabilization was also observed with longer cells (10 mm and 20 mm), however thin cells help enhance dispersive effects over absorptive ones.
\bibitem{DF}Our home-made external cavity is tuned thanks to a low-voltage piezoelectric actuator (displacement of about 6$\mu$m/volt).
\end{thebibliography}
\end{document}